\documentclass[aps,prx,twocolumn,floats,showpacs,superscriptaddress,nofootinbib]{revtex4-1}
\usepackage{graphicx,epsfig}
\usepackage{times,bbm}
\usepackage{graphics,dcolumn,bm,float}
\usepackage{amssymb,amsmath,rotate,color}
\usepackage[title,titletoc,toc]{appendix}
\usepackage{mathtools}
\usepackage{booktabs}
\usepackage{tcolorbox}
\usepackage[pagebackref=false,colorlinks,linkcolor=magenta,citecolor=blue,urlcolor=magenta]{hyperref}

\usepackage{wrapfig}
\usepackage{lipsum}
\usepackage{mwe}
\usepackage[mathlines]{lineno}
\usepackage{hyperref}
\usepackage{breqn}
\usepackage{bbold}
\usepackage{pgfplots}
\usepackage{float}
\usepackage{tkz-euclide}
\usepackage{braket}
\usepackage{physics}
\usepackage[caption=false]{subfig}
\usepackage[export]{adjustbox}
\usepackage{tikz}
\usetikzlibrary{through,calc}
\usetikzlibrary{positioning}

\newcommand{\be}{\begin{equation}}
\newcommand{\ee}{\end{equation}}

\newcommand{\bearr}{\begin{eqnarray}}
\newcommand{\eearr}{\end{eqnarray}}

\newcommand{\eps}{\varepsilon}

\newcommand{\up}{\uparrow}

\newcommand{\dn}{\downarrow}

\newcommand{\bq}{{\boldsymbol{q}}}

\newcommand{\br}{{\boldsymbol{r}}}

\newcommand{\bs}{\boldsymbol}
\newcommand{\bmt}{\left[\begin{matrix}}
\newcommand{\emt}{\end{matrix}\right]}

\newcommand{\bsq}{{\boldsymbol {q}}}
\newcommand{\bsk}{{\boldsymbol {k}}}

\begin{document}
\preprint{}
\title{Sound of Fermi arcs}

\author{F. Adinehvand}
\affiliation{Department of Physics$,$ Sharif University of  Technology$,$ Tehran 11155-9161$,$ Iran }
\author{Z. Faraei}
\affiliation{Department of Physics$,$ Sharif University of  Technology$,$ Tehran 11155-9161$,$ Iran }
\author{T. Farajollahpour}
\affiliation{Department of Physics$,$ Sharif University of  Technology$,$ Tehran 11155-9161$,$ Iran }
\author{S.A. Jafari}
\email{jafari@physics.sharif.edu}
\affiliation{Department of Physics$,$ Sharif University of  Technology$,$ Tehran 11155-9161$,$ Iran }

\date{\today}

\begin{abstract}
Using Green's function of semi-infinite Weyl semimetals we find that the collective charge excitations of
Fermi arcs in undoped Weyl semi-metals are linearly dispersing gapless plasmon modes. 
The gaplessness comes from proper consideration of
the deep penetration of surface states at the end of Fermi arcs into the interior of Weyl semimetal. 
The linear dispersion -- rather than square root dispersion of {\em pure 2D} electron systems with extended Fermi surface -- arises from the
strong anisotropy introduced by the Fermi arc itself, due to which the continuum of surface particle-hole excitations
in this system will have strong similarities to the one-dimensional electron systems. This places the Fermi arc
electron liquid in between the 1D and 2D electron liquids. Contamination with
the particle-hole excitations of the bulk gives rise to the damping of the Fermi arc sounds.
\end{abstract}



\maketitle
\narrowtext

\section{Introduction}\label{Sec1} 
The excitation spectrum of electronic systems in atomic length scales is composed of
individual electrons, while beyond a certain length scale, their collective character
becomes manifest~\cite{Phillips,PinesBook,BohmPines1953}. Among the salient collective excitations in electron liquids
are plasmon which are longitudinal collective density oscillations~\cite{PinesBook}. 
Based on general hydrodynamic arguments, plasmons of three dimensional (3D) systems 
are always gapped and given by $\omega_{\rm{p,3D}}^2\sim ne^2/\epsilon_0 m$~\cite{kittel1976,Phillips}.
In two-dimensions, both quantum mechanical~\cite{vignale} and hydrodynamic arguments~\cite{fetter1973fetter} 
establish that the long wave length limit of the plasmon dispersion must be $\propto \sqrt q$. 
This holds in both Dirac materials~\cite{hwang2009plasmon,wunsch2006dynamical} and 2D materials with quadratic 
band dispersion~\cite{vignale}. 
Plasmons in a two-dimensional electron gas were first observed for a sheet of electrons on liquid helium~\cite{grimes1976plasmons,grimes1976observation} 
and then in inversion layers~\cite{allen1977observation}, as well as in graphene~\cite{bostwick2007quasiparticle}.

\begin{figure}[b]
 \includegraphics[width=4cm]{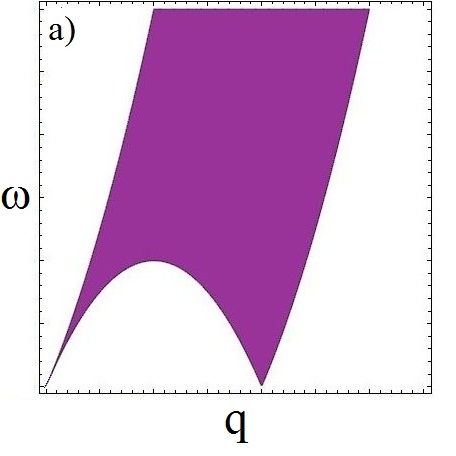}
  \includegraphics[width=4cm]{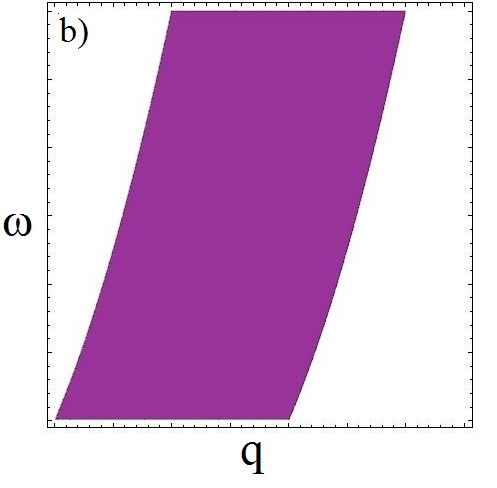}
   \includegraphics[width=5cm]{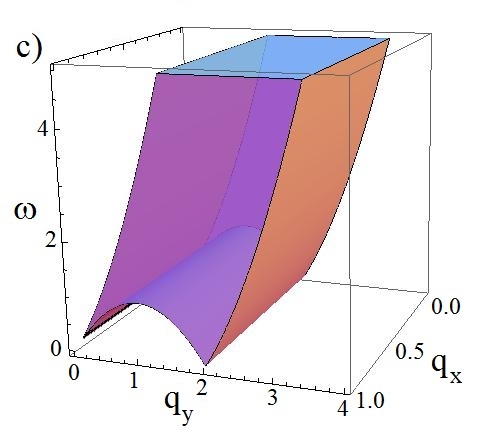}
  \caption{Particle-hole continuum in (a) 1D, (b) 2D and (c) Fermi arc electron liquid. 
  In panel (b) the PHC of the 2D (as well as 3D) electron liquid is obtained 
  from the rotation of the shaded area around the energy axis. 
}
    \label{PHC.fig}
     \end{figure}

On the other hand, in one-dimension (1D) the fixed point of interacting gapless systems is the Luttinger liquid~\cite{voit1995one}, 
rather than a Fermi liquid~\cite{BaymPethick}.  
The low-energy sector of the spectrum of interacting fermion systems in 1D 
is exhausted by collective excitations only. In 1D, the collective charge density excitations appear in the form of
linearly dispersing sound waves~\cite{senechal2004,gogolin2004bosonization,rao2001bosonization}. 
The intuitive picture for collective charge and spin excitations in 1D is based on Fig.~\ref{PHC.fig}-a. 
At low energies and low momenta, the particle-hole (PH) continuum (PHC) is composed of a narrow band of 
free PH excitations all moving at the same group velocity, such that any amount of Coulomb forces binds them
into coherent collective modes~\cite{GiamarchiBook}. This is in contrast to the PHC of higher dimensional
electron gases with extended Fermi surface in Fig.~\ref{PHC.fig}-b~\cite{vignale} where due to rotational
degree of freedom of the wave  vector $\bsq$ there is no
coherence in the group velocity of the particle-hole pairs. 
Therefore the electron liquids in dimensions higher than one, do not sustain plasmonic sound waves, except
for a genuine non-equilibrium situation recently proposed for the bulk of 3D Weyl semimetals~\cite{song2019hear}. 
In this work we will show that the electron gas formed by Fermi arc states on the surface of Weyl semimetals (WSMs)
supports a linearly dispersing gapless plasmon mode. 


In general, restricting the material in half space gives rise to surface plasmons which were employed by Ritchie to
explain the energy loss of fast electrons passing through thin films~\cite{ritchie1966surface}. 
Since the penetration depth of electromagnetic waves in metals is rather short, the
plasmons of metal surfaces are major players in technical applications of the collective 
oscillations~\cite{maier2005plasmonics,moore1965cramming,ozbay2006plasmonics,atwater2007promise,heber2009plasmonics}.
In recent year, Weyl semimetals have been added to the list of conducting materials~\cite{armitage2018weyl,rao2016weyl,yan2017topological}
which are additionally endowed with topological indices protecting their gapless nature. 
The bulk electronic states in these systems is composed of linearly dispersing bands
that touch each other at nodal points~\cite{lu2015experimental,huang2015weyl}. WSMs are characterized by very peculiar
surface states known as Fermi arc~\cite{xu2016observation} which are localized on the
surface if the Fermi wave vector is in the middle of the arc, but penetrate deeply into the bulk as 
the (Fermi) wave vector approaches the two ends of the Fermi arc~\cite{faraei2018green}. 

Bulk plasmons and associated 
surface plasmon resonance (resulting from the projection of dielectric function on the surface)
has been studied in WSMs~\cite{lv2013dielectric,andolina2018quantum}. 
In these systems the gap in the 3D plasmons is given by $\sim e^2 n^{2/3}v_F$. 
Assumption of specular reflection of incident electrons 
is equivalent to projection of the bulk dielectric function to the $z=0$ boundary surface.
Using this approach, Lo\v{s}i\'c obtained gapped surface plasmon resonances for doped Weyl semi-metals~\cite{lovsic2018surface}. 
Employing a simple dispersion of Fermi arcs, and neglecting the associated matrix elements, 
Lo\v si\'c finds a plasmon mode dispersing as $\sqrt{q_y}=\sqrt{q\sin\theta}$ ($y$ direction is perpendicular to the Fermi arc). 
Apart from the angular dependence, the $\sqrt q$ behavior is expected for a {\it pure} 2D system with {\em extended Fermi surface}.
Andolina and coworkers have studied the same problem~\cite{andolina2018quantum}. 
If only the contribution of arc states is
taken into account, the plasmons for the Fermi arc states in their work becomes gapped.

 
\begin{figure}[t]
 \includegraphics[width=7cm]{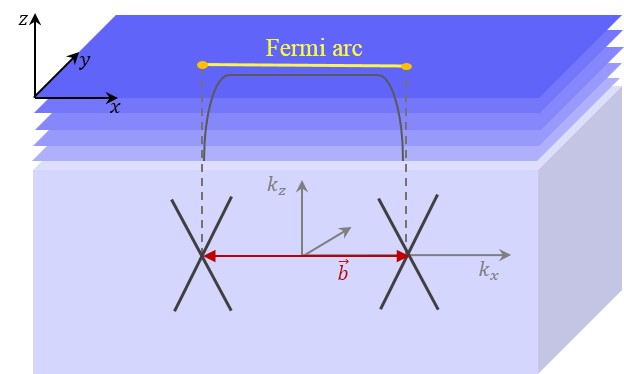}
  \caption{ Schematic demonstration of the Fermi arc on the sufrace of a Weyl semimetal that connects the
  projection of two Weyl nodes on the surface. The localization lenghth of Fermi arc states increases by 
  approaching the projection of Weyl nodes. This has been shown by the gray line.}
    \label{sample.fig}
     \end{figure} 

In this paper we use our previously developed Green's function for WMSs~\cite{faraei2018green} which allows us
to isolate the effects solely arising from the Fermi arcs states.
Note that our Green's function approach does not assume any effective Hamiltonian to describe the
Fermi arc states. Fermi arc states in our approach are faithfully produced from appropriate boundary conditions~\cite{faraei2018green}.
Then we find that the deep penetration of end-of-Fermi-arc states into the bulk, 
combined with the peculiar form of PHC -- which bears certain resemblance to the 1D case as in Fig.~\ref{PHC.fig}-c --
will compromise to produce a gapless branch of linearly dispersing Fermi arc plasmons. Let us briefly elaborate on
these points: (i) As can be seen in Fig.~\ref{sample.fig}, the penetration length of Fermi arc states
into the interior of the WSM increases more and more by approaching the two ends of Fermi arcs. The method
employed in Ref~\cite{andolina2018quantum} uses a projection of the polarization function onto the surface, 
and avoids dealing with a determinantal equation in real-space coordinates $z,z'$. Evading the determinant
misses the long tail of Fermi arc states. We will show how systematic
treatment of the penetration effects gives rise to gapless plasmon mode by progressively increasing the 
size of the determinant involved. 
(ii) The Fermi arc per se, creates enormous anisotropy
in such a way that the phase space for the scattering of electrons will resemble those of
one-dimensional systems as depicted in Fig.~\ref{PHC.fig}-c. This is in sharp contrast
to the systems with extended closed Fermi surface where a rotational degree of freedom
of the wave vector $\bsq$ of low-energy scattering processes gives rise to a PHC of type (b)
rather than type (c) in Fig.~\ref{PHC.fig}. 
In this way the resulting plasmon mode
will become a (linearly dispersing) sound wave. Note that such a Fermi arc sound wave is entirely different from the 
chiral zero sound (CZS) considered in Ref.~\cite{song2019hear} in the following respects: 
(i) the most important difference is that to generate the CZS one needs to generate a {\em non-equilibrium} 
population around the right and left Weyl nodes. 
(ii) to realize the CZS the inter-valley relaxation rate must be much smaller than intra-valley time scales. 
(iii) realization of {\em chiral} zero sound requires a background B field to generate right/left movers, i.e. it is built on the chiral anomaly.

\section{Model and formalism}

We consider a semi-infinite Weyl semimetal in $ z<0 $ part of the space bounded by $ z=0 $ surface as illustrated in Fig.~\ref{sample.fig}.
In our previous works~\cite{faraei2018green,FaraeiScWeyl} we have used the Green's function approach to formulate the Fermi arcs.
We do not {\em assume} any particular Hamiltonian for the Fermi arc states. We rather obtain the Fermi arcs from appropriate boundary 
conditions~\cite{faraei2018green}. In this way, we ensure that all the necessary matrix elements, as well as the tail of 
Fermi arc states in the bulk (see Fig.~\ref{sample.fig}) are properly taken into account. 
The deep penetration of the states near the ends of Fermi arc into the interior of WSM is one of the essential
differences of the electron gas formed by the Fermi arc with respect to the standard 2D electron gases. 
Missing this effect will give rise to $\sqrt q$ behavior~\cite{lovsic2018surface} of the normal 2D electron gas systems.

When the translational symmetry along the $z$ axis is broken by presence of a surface at $z=0$, the dielectric
function will become a matrix in spatial $z,z'$ indices. 
The "dielectric matrix" in the random phase approximation is given by,
\be
 \eps_{z,z'} (q,\omega)=1-V_{z,z'}(q)\Pi_{z,z'}(q,\omega),
 \label{epsilon}
\ee 
where $ V_{z,z'}(\vec q)=2\pi e^2 e^{-q|z-z'|}/q$ is the Coulomb matrix element with $\vec q=(q_x,q_y)$ denoting
the wave vector in the plane. Here $z$ and $z'$ are the distance of the two electrons from the boundary. When they 
are at the same plane, $z=z'$, this reduces to the Coulomb matrix element of the 2D systems~\cite{Katsnelsonbook}. 
$\Pi_{z,z'} (\vec q,\omega)$ is the polarizability tensor which can be written in terms of the
Green's function as,
\be
\Pi_{z,z'} (\vec q,\omega)= \sum_{\vec k, \nu} G_{z,z'}(\vec k,\omega) G_{z',z}(\vec k +\vec q,\omega+\nu).
\label{pi.eqn}
\ee 
From this point we do not perform any further approximations such as projection of the polarization function into the
surface~\cite{lovsic2018surface} or approximating the determinants such as ${\det}~\eps_{z,z'}=0$ by related trace operations~\cite{andolina2018quantum}.
Instead, following Andersen and coworkers~\cite{andersen2012spatially} the plasmon resonance in a generic situation with broken
translation invariance are obtained numerically. 
For this purpose, one needs to numerically solve the eigenvalue equation,
\be
\int \eps(\br,\br',\omega)\phi_n(\br',\omega)d\br'=\varepsilon_n(\omega)\phi_n(\br,\omega),
\label{denmark.eqn}
\ee
and then to find zero eigenvalues $\varepsilon_n(\omega)=0$. 
By scanning range of $\omega$ values, one can find the energy scales $\omega_p$ at which the eigenvalues of the dielectric matrix vanish, $\eps_n(\omega_p)=0$.
This will correspond to plasmon resonances. 
Various branches are labeled by $n$. We are interested in the lowest energy branch. 
Furthermore at a resonance the loss function $ \Im(\varepsilon_n^{-1}) $ 
will develop a peak which corresponds to the characteristic energy losses suffered by fast charged particles 
traversing the material~\cite{winther2015quantum}. To take the long penetration
depth of Fermi arc states into account, one needs to discretize the portion of space near the surface as depicted in Fig.~\ref{sample.fig}. 
The separation between the Weyl nodes is $2|\vec b|$, where $b$ can be used as a unit of momentum, which will also set the unit of length as $\hbar/b$. 

To calculate the $\eps(\br,\br',\omega)$, one basically needs the polarization function $\Pi(\br,\br',\omega)$ 
as they are related by Eq.~\eqref{epsilon}. In the geometry of Fig.~\ref{sample.fig} by translational symmetry in 
$(x,y)$ plane, the polarization will reduce to $\Pi(z,z',\bq,\omega)$ where $\bq=(q_x,q_y)$
is the wave vector corresponding to the surface. To calculate the polarization we can use the  
Green's functions obtained for the Weyl Hamiltonian by applying a proper boundary condition~\cite{faraei2018green}. 
As we are interested in bare Fermi arc plasmons, we use the particular surface Green's functions that 
are naturally separated from the bulk contribution. 
Without loss the generality, we rotate the coordinate system, such that the Fermi arc is oriented 
on the $k_x$ axis. In this situation the Green's function will become much easier to work with~\cite{FaraeiScWeyl}. 
Convolution of the Green's functions (which are $2\times 2$ matrices in their spin indices) as in Eq.\eqref{pi.eqn} gives,
\begin{widetext}
\be
 \begin{aligned}
\Pi_{z,z'}^{\uparrow\uparrow(\downarrow\downarrow)} =& \dfrac{e^{-b\zeta}q_y}{4\pi^4\zeta^3(\omega- q_y)} 
\bigg\lbrace  \sinh (q_x\zeta/2)
\big[ b\zeta^2\cosh b\zeta-(\zeta+b\zeta^2) \sinh b\zeta \big] (2q_x \mp iq_y)\\
&+ \cosh (q_x\zeta/2) 
\big[ (4b^2 \zeta^2 +4+4b\zeta \mp iq_xq_y \zeta^2)\sinh \zeta b-(4b\zeta+4b^2 \zeta^2)\cosh b\zeta \big] \bigg\rbrace ,\\
\Pi_{z,z'}^{\uparrow\downarrow(\downarrow\uparrow)} =&
\dfrac{e^{-b\zeta}q_y}{4\pi^4\zeta^3(\omega- q_y)} 
\bigg\lbrace  \sinh (q_x\zeta/2)
\big[ (q_xq_y\zeta^2\mp 4i\zeta b\mp 4i\zeta^2b^2\mp 4i)\sinh \zeta b\pm(4i\zeta^2b^2+4i\zeta b)\cosh b\zeta \big]\\
&+ \cosh (q_x\zeta/2)\big[ -(\zeta +b\zeta^2)\sinh \zeta b+b\zeta^2\cosh \zeta b \big](q_y \mp 2i q_x) \bigg\rbrace,
\end{aligned}
\label{pi2}
\ee
\end{widetext}
where $2\vec b=(2b,0,0)$ is the vector that connects the Weyl nods in the bulk of the WSM and $\zeta=2(z+z')$ with 
$z$ and $z'$ measured from the surface. The dispersion of Fermi arc states involves single-particles moving along
positive $y$ direction on the top surface. Therefore PH excitations with positive energy are restricted
to $q_y>0$ on the top surface. The $q_y<0$ PH excitations correspondingly take place in the bottom surface. 
A nice property of the above polarization matrix elements is that under mirror reflection with respect to Fermi arc, $\varphi\to -\varphi$
or $q_y\to -q_y$ (which is equivalent to going from top surface to bottom surface) we have
\be
  \Pi^{\alpha\beta}(q,\varphi,\omega)\to \alpha\beta \Pi^{\bar\alpha\bar\beta}(q,-\varphi,-\omega)
  \label{inversion.eqn}
\ee
where $\alpha,\beta$ can take $\up\equiv+1$ or $\dn\equiv -1$ values. 

\begin{figure}[t]
  \includegraphics[width=4.2cm]{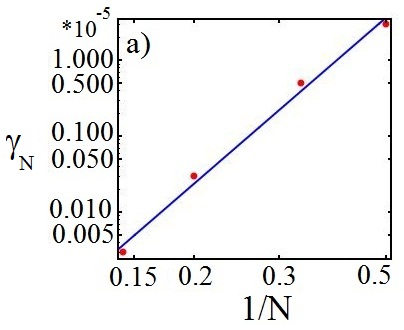}
  \includegraphics[width=4.1cm]{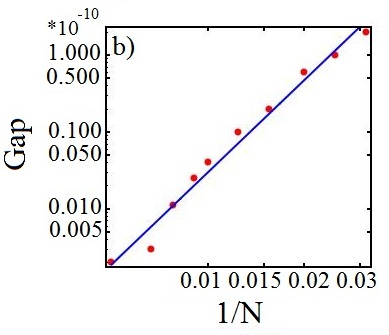}
   \caption{Finite size scaling analysis of the relative slope difference $\gamma_N$ (left)
   and the gap $\Delta_N$ (right) of the lowest Fermi arc plasmon modes. Upon increasing $N$
   the gap quickly vanishes and the slopes quickly become degenerate.
   }
     \label{convergence.fig}
      \end{figure}

To solve for $\det\eps=0$, 
we need to discretize the $z$ direction in Eq.~\eqref{pi2} and form the $\Pi^{\alpha,\beta}_{z,z'}$ matrix. Discretization by a
mesh of size $N$, we will be dealing with $2N\times 2N$ matrices, where the extra $2$ comes from
the above matrix structure in the spin space. The numerical details are given in the supplement. 
To get a feeling about the Fermi arc plasmon dispersion,
let us start with the $N=1$ approximation which can be analytically handled. In this case, 
we just need to consider one layer at $ z=0 $. 
Eq~\eqref{epsilon} for one layer gives,
\be 
\eps=\begin{pmatrix}
1-V_0\Pi^{\uparrow \uparrow} &-V_0 \Pi^{\uparrow \downarrow}\\
-V_0 \Pi^{\downarrow \uparrow}& 1-V_0\Pi^{\downarrow \downarrow}
\end{pmatrix}
\ee
where $ V_0=V_{z=0,z'=0}(q)=2\pi e^2/q $ is pure 2D Coulomb interaction matrix element. 
In $\bsq\to 0$ limit, the determinant of the above dielectric matrix has the following solution which 
after restoring various constants gives the following $q$-linear plasmon dispersion,
\be
\hbar\omega= \hbar q v_F\sin\varphi (1\pm \gamma_1)+ \Delta_1 
\label{plasmonN1.eqn}
\ee
where $\gamma_1=\alpha_{\rm WSM}\sqrt{\sin^2\varphi+4\cos^2\varphi}/(2\pi^2)$ is the relative difference
in the slope of two modes and is determined by the "fine structure" constant $\alpha_{\rm WSM}=e^2/(\hbar v_F)$ 
which is of the order of unity for Fermi velocities that are hundreds of times less than the speed of light. 
The gap for $N=1$ approximation is given by $\Delta_1=\frac{2e^2b}{\pi^3}\sin\varphi $. 
But the two-dimensional electron systems are not expected to have gapped plasmons~\cite{vignale,Katsnelsonbook}. 
In fact anticipating that $\gamma,\Delta$ will depend on the mesh size $N$, we have
introduced subscript $"1"$ in $\gamma_1$ and $\Delta_1$. 
To investigate this systematically, in Fig.~\ref{convergence.fig}
we study the effect of taking progressively deeper layers around the surface into account. 
As can be seen, by increasing $N$ both $\gamma_N$ (left panel) and $\Delta_N$ (right panel) approach zero. 
This figure clearly shows that the gap $\Delta_{N=1}$ and the slope difference $\gamma_{N=1}$ are
artifacts of considering a mesh of size $N=1$ and quickly vanish by taking progressively larger mesh
sizes into account. 
As pointed out, this form of finite size effect in surface states arises because the Fermi arc states
at the two ends of the Fermi arc deeply penetrate into the bulk as in Fig.~\ref{sample.fig}. 
The conclusion is that the numerical evaluation of the determinant of $\eps$ for reasonably large $N$ can not be avoided.
Evading this determinant by projecting it to the surface -- which might be valid for 
typical non-Weyl electronic systems -- introduces sever errors in the dispersion of Fermi arc plasmons~\cite{lovsic2018surface,andolina2018quantum}. 

\begin{figure}[b]
 \includegraphics[width=9cm]{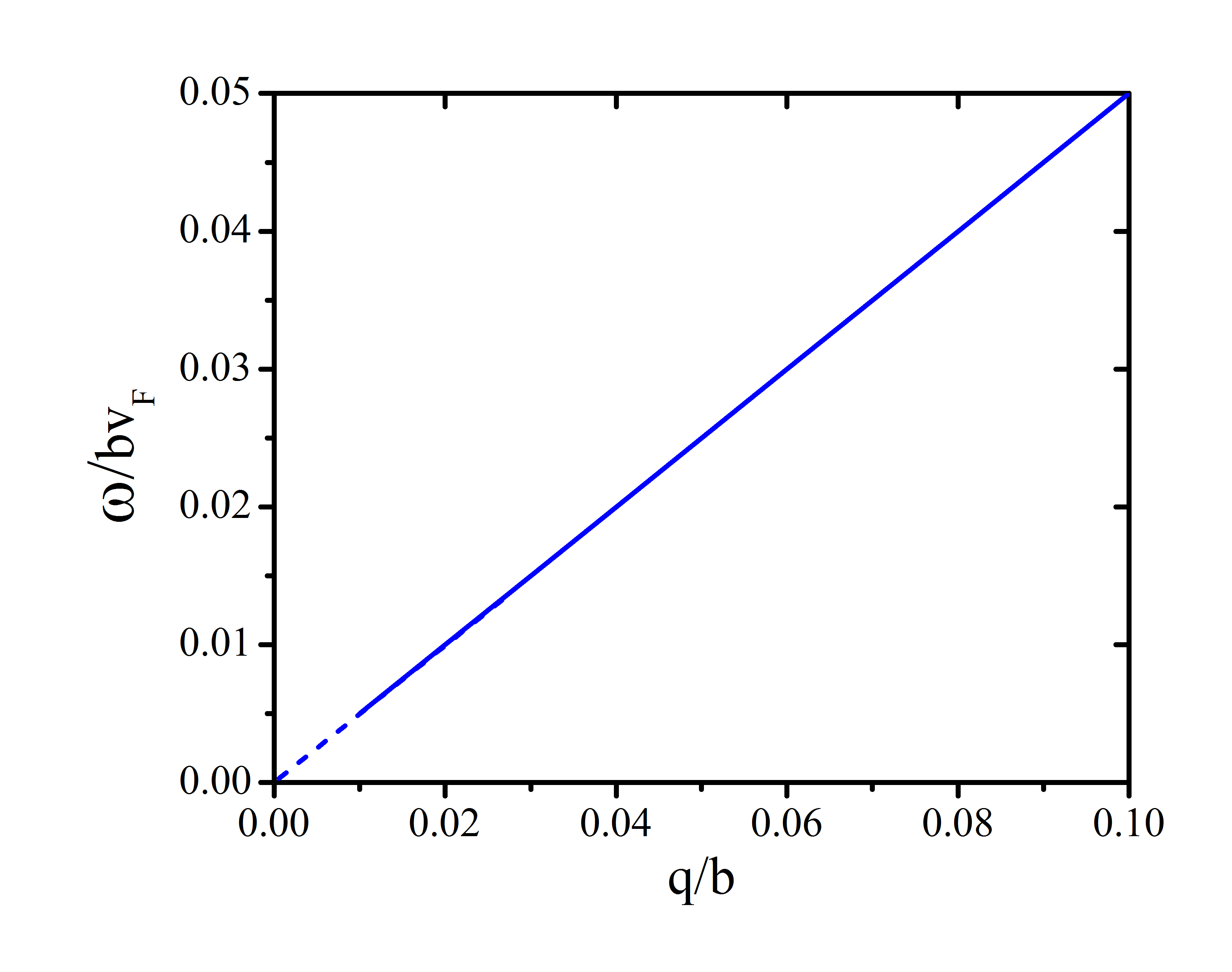}
  \caption{Lowest surface plasmon mode of Fermi arc states for a grid of $N=50$ layers
  with penetration depth of $db=25$. The polar angle is fixed by $\varphi=\pi/6 $.}
    \label{w-q.fig}
     \end{figure} 
Although the $N=1$ approximation to determinant of $\eps$ does not give correct
gap and slope, but the linear nature of the plasmon dispersion and its angular 
dependence survive the larger $N$ limit. To establish the linear $q$ dependence and
$\sin\varphi$ angular dependence, in Fig.~\ref{w-q.fig}
we have plotted the dispersion of the (doubly degenerate) lowest plasmon mode for large enough $N=50$ to ensure
that $\gamma_N$ and $\Delta_N$ are already zero. 
The direction of the wave vector $\bsq$
is fixed by $\varphi=\pi/6$ where $\varphi=\arctan\big(q_y/q_x \big)$ is the polar angle
of $\bsq$. Note that in this figure both $\omega$ and $\bsq$ are plotted in their
natural units. As can be seen the linear dispersion of the $N=1$ approximation, 
surprisingly robust at much larger $N$. 
Extrapolating the dispersion to $q\to 0$ limit clearly confirms that for this value of mesh 
size, the plasmon mode is gapless. Also the fact that in Fig.~\ref{convergence.fig} the 
relative slope difference $\gamma_N$ is already zero for $N=50$ indicates that this
lowest mode is doubly degenerate. The double degeneracy corresponds to $\up$ and $\dn$ spin
directions. 

\begin{figure}[t] 
 \includegraphics[width=8cm]{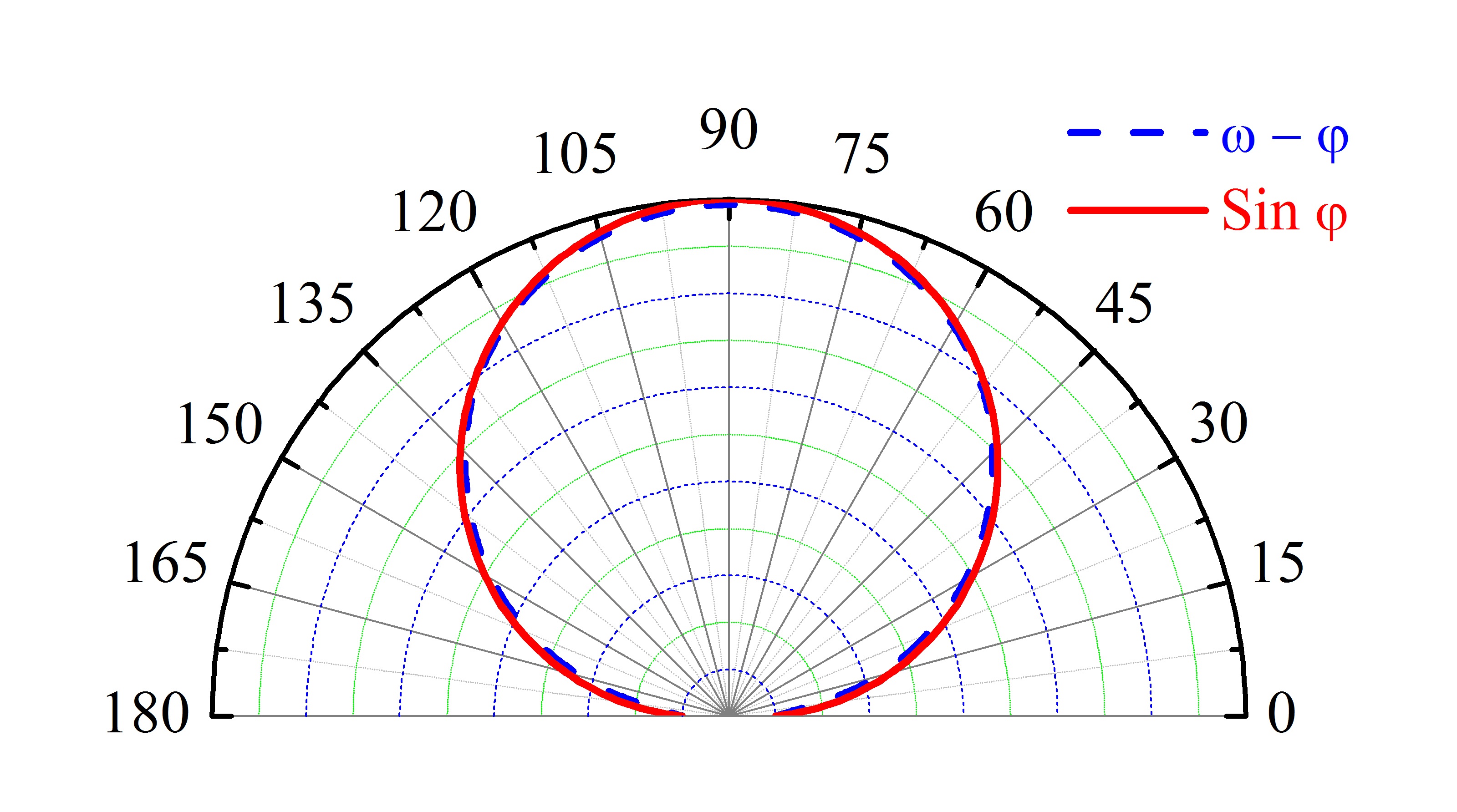}
  \caption{The lowest surface plasmon frequency (dashed blue) and $ \sin\varphi $ function
   (solid red) as a function of the polar angle $\varphi$ for $N=50$ and $db=25$ for a fixed $q=0.01b$ wave vector magnitude. }
     \label{wphi.fig}
      \end{figure}

Let us now investigate the angular dependence of the linear plasmon mode of the Fermi arcs. 
The slope of the lowest Fermi arc plasmon mode in Fig.~\ref{w-q.fig} within the machine precision is given by $\sin(\pi/6)=0.5$. 
To explore this further, in Fig.~\ref{wphi.fig} we have plotted the lowest mode plasmon frequency 
as a function of the polar angle $ \varphi $ for a fixed $|\bsq|=0.01b$. The magnitude of wave vector
has been chosen small enough to ensure that it already lies in the linear dispersion regime. 
The dashed line denotes the numerical data, while the solid line represents the $\sin\varphi$ profile. 
It is surprising to see that the $\sin\varphi$ angular dependence also robust even in the limit
of larger mesh sizes. Therefore properly handling the penetration of Fermi arc states
into the interior of WSM gives rise to a branch of linear plasmon mode which is
doubly (spin) degenerate and disperses as $\omega=v_F q\sin\varphi=v_F q_y$. 

\begin{figure} [t]
  \includegraphics[width=8cm]{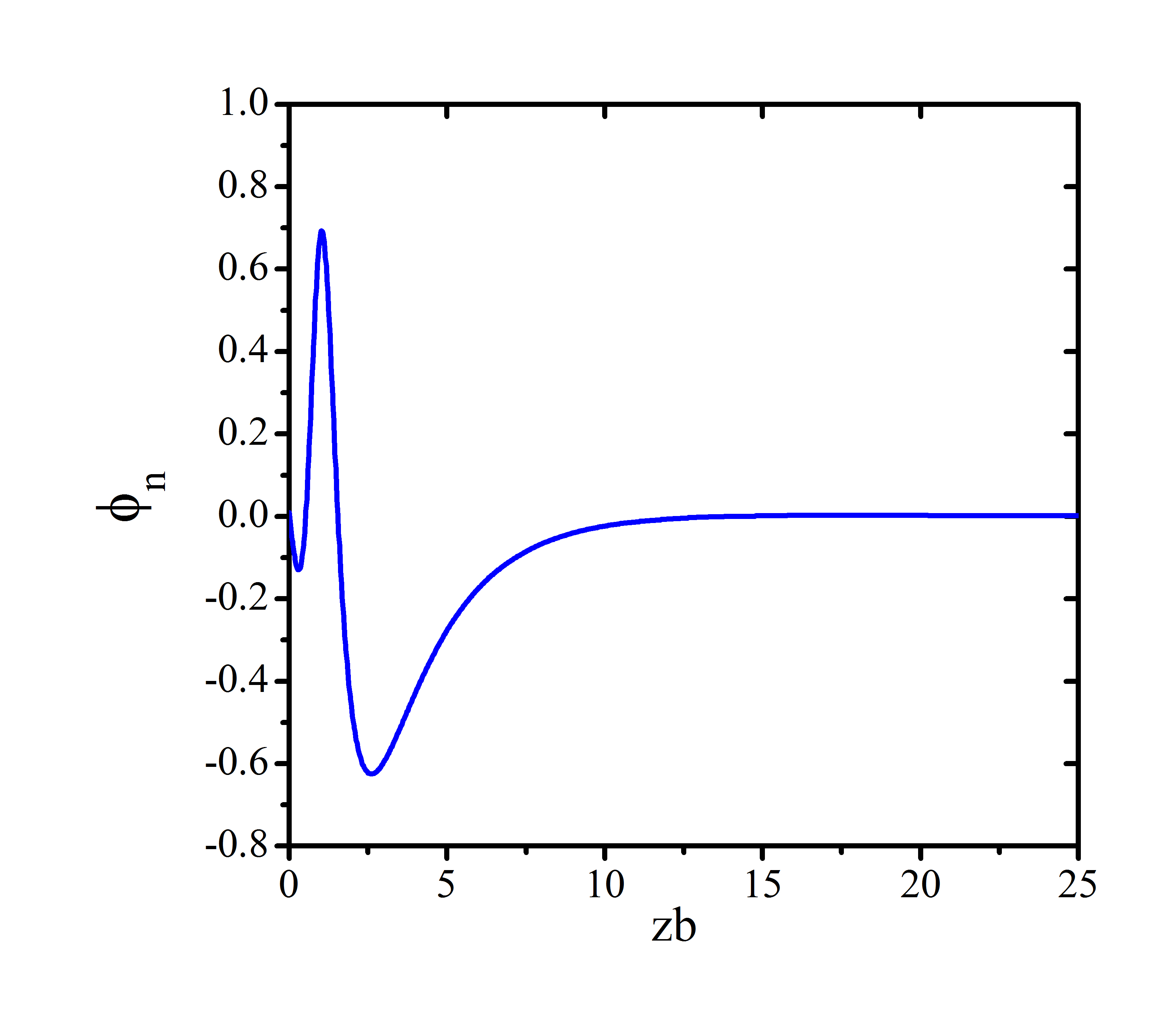}
   \caption{The electrostatic potential profile for spin $\up$ and $\dn$ (degenerate) Fermi arc electrons 
    as a function of distance $zb$ from the surface for $ \varphi=\pi/6 $, $|q|=0.1$ and $N=150$. 
    The modes are localized on the surface.}
    \label{wave}
    \end{figure}

Finally in Fig.~\ref{wave} we have plotted the eigen-functions of Eq.~\eqref{denmark.eqn}
that correspond to zeros of the dielectric function $\eps_n(\omega)$. The eigen-function
corresponds to lowest mode evaluated for parameters $N=150$ and $db=75$ and $q=0.1b$. For such 
a large value of $N$ the slope difference $\gamma_N$ is already zero, and this electrostatic
profile is doubly degenerate corresponding to $\up$ and $\dn$ spin densities. 
The actual density of electrons according to Poisson equation will be the second
derivative of this curve. The mode is clearly bound to the surface of the WSM.

\section{Fermi arc as a unique electron gas}
Typically for an extended Fermi surface in 2D, one expects a $\sqrt q$ plasmon dispersion.
But the linear plasmon $q\sin\varphi=q_y$ arising from the Fermi arc states (of undoped WSMs)
is in sharp contrast to the usual 2D plasmons. Even more surprising is that this behavior
does not receive any correction by numerically computing the plasmons for larger $N$. So 
there must be a reason for this behavior. To elaborate on this, let us start by 
noting that apart from diploar $\sin\varphi$ angular dependence (which is a 2D feature),
the linear sound mode is reminiscent of the 1D bosonic modes. 
Indeed due to strong anisotropy of the Fermi arc, the PHC of Fermi arc states in 
Fig.~\ref{PHC.fig}-(c) bears more resemblance to the PHC of 1D systems in panel (a), rather than the PHC of 2D systems in panel (b). 

If the dispersion $\varepsilon(k_y)=k_y$ of the Fermi arc was a 1D dispersion,
then a straightforward bosonization arguments~\cite{gogolin2004bosonization} would give
the charge bosonic collective mode with $\omega(q_y)\propto q_y$ dispersion. 
The dispersion of Fermi arc states is quite similar, $\eps^{\rm arc}(\bsk)=k_y$, albeit with the importance
difference that $k_y$ is now a component of a 2D vector $\bsk=(k_x,k_y)$. 
The restricted phase space for the bosonic mode formation in Fig.~\ref{PHC.fig}-(c) within 
the random phase approximation gives the bosonic mode $\omega(\bsq)\sim q_y=|\bsq|\sin\varphi$. 
The surprising fit of the numerical data obtained from RPA dielectric matrix suggest that the anisotropy
of the Fermi arc states is likely to admit interesting forms of bosonization which may allow one
to go far beyond the above simple RPA treatment. From this point of view, the electron gas
formed by the Fermi arc states is actually something between 1D and 2D.

The peculiar 1D-like continuum, the Fermi arc plasmons will protect them from
Landau damping into the surface PH excitations. However, contamination with the
bulk PHC of excitations which is expected to be stronger for those Fermi arc states
closer to the ends of Fermi arc can lead to the damping of Fermi arc plasmon sounds.

\section{Summary and discussion}
Within the random phase approximation, 
we studied the collective charge dynamics of the Fermi arc states of undoped Weyl semimetals. Fermi arc states near the projection of
Weyl nodes have infinite localization lengths. As such, their tail extends well inside the bulk (see Fig.~\ref{sample.fig}). 
The proper treatment of these tails by considering large enough grids makes the collective excitations gapless and the $\up$ and $\dn$ 
modes degenerate. 
Failing to account for this long tail, will generate gapped plasmons~\cite{lv2013dielectric,andolina2018quantum} from pure Fermi arc states. 
The peculiar dispersion of Fermi arc states places them somewhere in between the 1D and 2D electron liquids (Fig.~\ref{PHC.fig}).
As such, unlike the typical 2D electron liquids possessing extended Fermi surface (and hence a finite matter density) 
for which the plasmon disperses as $\sqrt q$,  for Fermi arcs we find
a plasmon branch dispersing like $q\sin\varphi$ where $\varphi$ is the polar angle of the wave-vector $\bsq$. The linear
dispersion is more like a 1D feature which arises from the limited phases space for particle-hole excitations in Fig.~\ref{PHC.fig}~(c). 
This peculiar PH continuum endows the Fermi arc states with a sound branch arising from charge oscillations of the electrons. 
The p-wave profile (i.e. $\sin\varphi$ dependence) is the essential difference of the present sound from similar sounds (bosons) in 1D electronic systems. 

Unlike the recent proposal for the chiral zero sound of bulk electronic degrees of freedom in Weyl semimetals~\cite{song2019hear},
the present sound is due to Fermi arcs states and moreover, does not require non-equilibrium population of the two Weyl nodes. 
Mixing of the Fermi arc states with bulk states gives rise to contamination of the bulk PHC that opens up a damping channel
for the Fermi arc sounds~\cite{andolina2018quantum}. 
\section{acknowledgments}
T.F. appreciates the financial support from Iran National Science Foundation (INSF) under post doctoral project no. 96015597. 
S. A. J. appreciates research deputy of Sharif University of Technology, grant no. G960214. Z.F. was supported by 
a post doctoral fellowship from the Iran Science Elites Federation (ISEF). 

\bibliographystyle{apsrev4-1}
\bibliography{mybib}

\newpage
\onecolumngrid

\appendix
\begin{center}
{\bf Supplementary material}
\end{center}
In this supplementary material we following Refs.~\cite{andersen2012spatially} and~\cite{winther2015quantum} of 
the main text, we summarize how to solve the plasmon eigenvalue problem in confined geometries. 
Longitudinal modes are zeros of the dielectric function. 
In generic situation corresponding to an arbitrary geometry, the dielectric function will be 
a function of two separate spatial coordinates $\bs r$ and $\bs r'$. The longitudinal modes arise 
from zeros of the dielectric function which can be viewed as a matrix in indices $\bs r$ and $\bs r'$. 

\begin{figure}[b]
 \includegraphics[width=7cm]{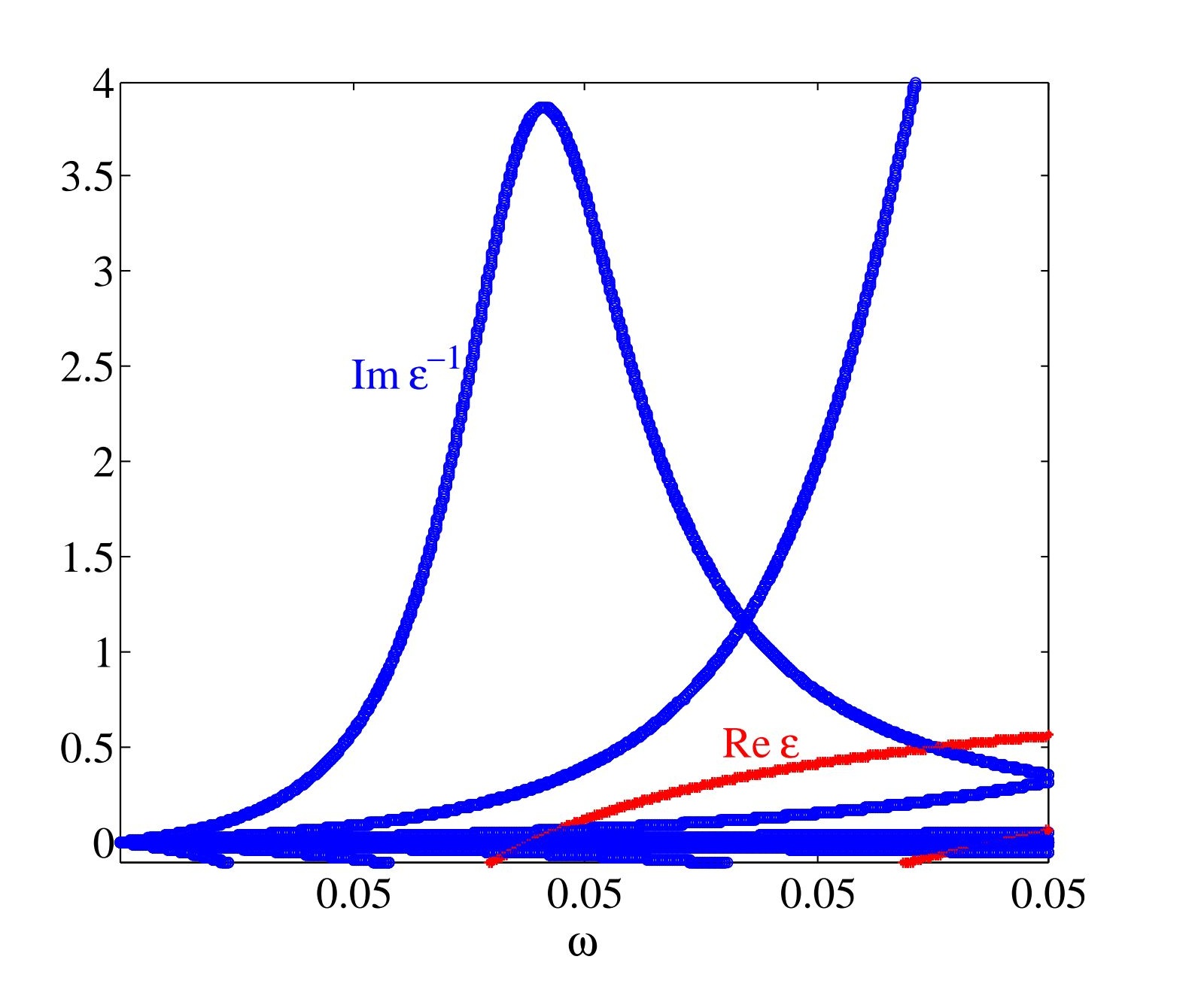}
  \caption{The red plot is the real part of the eigenvalues $\eps(\omega)$ of the dielectric
  matrix. The blue plot is the imaginary part of the loss function  
  $ \eps^{-1}(\omega)$ as a function of frequency $\omega$ for a mesh with $ N=50 $ layers and the
  horizontal momentum$ \bsq=0.1 $.}
    \label{e-w.fig}
     \end{figure}
One therefore needs to finds the zero modes of the following equation,
  \begin{equation}
    \int \eps({\bs r},{\bs r'} ,\omega) \phi_{n}({\bs r'},\omega)d{\bs r'}=\eps_n(\omega)
    \phi_{n}({\bs r}, \omega)
     \label{epsilon}.
      \end{equation}
where $n$ labels various modes, and we are searching for solutions where the eigenvalue $\eps_n(\omega)=0$.
Since we do not no apriori at which frequencies the dielectric egenvalues $\eps_n(\omega)$ become zero,
one has to scan a range of $\omega$ values to find where the real part of $\eps_n(\omega)$ crosses the zero. 
To this end a suitably chosen mesh can be used to descritize the matrix $\eps ({\bs r},{\bs r'},\omega)$ for a given fixed
$\omega$. Then for any given $\omega$, the eigenvalues of the above equation are obtained. Varying $\omega$ gives
typical plots like Fig.~\ref{e-w.fig}. 

For the present situation, the wave vector $\bsq$ in the $(q_x,q_y)$ plane is a good quantum number to lable the plasmons. 
However the translational invariance along $z$ is lost and therefore the dielectric matrix will be a matrix in
spatial indices $z,z'$ for given values of $\bsq$ and $\omega$. This can be used to generate Fig.~\ref{e-w.fig}
which are generated for a mesh of size $N=50$. 
The red plots are the real parts of $\eps_n(\omega)$ and the blue plots
are the imaginary parts of the loss function $1/\eps_n(\omega)$ measured in electron energy loss spectroscopy. 
Various plots correspond to various values of $n$. 
As can be seen in Fig.~\ref{e-w.fig} the plasmon resonance is a frequency $\omega_p$ at which
the real part of $\eps_n(\omega)$ vanishes {\em and} the loss function $\Im\left(\eps^{-1}_n(\omega)\right)$ developes a peak. 

Repeating the above procedure for all other values of $\bsq$, allows us to map the disperson of longitudinal (plasmon)
modes. We are interested in the lowest $n=0$ mode which are confined to the surface.

\end{document}